\documentclass[aps,prd,preprint,showpacs,floatfix,nofootinbib,superscriptaddress]{revtex4-1}

\usepackage{dcolumn}% Align table columns on decimal point
\usepackage{amsmath, amsfonts, amssymb, mathrsfs, times, float, color, bm}
\usepackage{extarrows}
\usepackage{graphicx}
\usepackage{graphicx,epstopdf}% Include figure files
\usepackage{bm}% bold math

\usepackage{exscale}
\usepackage{relsize}

\begin{document}

%\title[]{Coordinate transformation, gauge invariance and covariance in general relativity}

%\title[]{Covariance, gauge conditions and coordinate transformations in general relativity}
%\title[]{The weak-field limit for the metric for Kerr black hole in the radiation gauge to the first-order approximation}
%\title[]{The weak-field limit for the metric of Kerr black hole in radiation gauge}
\title[]{The weak-field-limit solution for Kerr black hole in radiation gauge}

\author{Chunhua Jiang}%
\author{Chaohong Pan}
%\email{jiangchunhua@usc.edu.cn}
\affiliation{School of Mathematics and Physics, University of South China, Hengyang, 421001, China}
\author{Wenbin Lin}
\email{To whom all correspondence should be addressed (email: lwb@usc.edu.cn).}
\affiliation{School of Mathematics and Physics, University of South China, Hengyang, 421001, China}
\affiliation{School of Physical Science and Technology, Southwest Jiaotong University, Chengdu, 610031, China}
\date{\today}

%\begin{abstract}
%Recently Chen and Zhu propose a true radiation gauge for gravity [Phys. Rev. D 83, 061501(R) (2011)]. This work presents a weak-field-limit solution for Kerr black hole in this radiation gauge.
%\keywords{radiation gauge \and Kerr black hole \and gravitational wave radiation}
%% \PACS{PACS code1 \and PACS code2 \and more}
%%\PACS{04.20.Cv \and 04.25.Nx \and 04.70.Bw}
%% \subclass{MSC code1 \and MSC code2 \and more}
%\end{abstract}

\begin{abstract}
{\color{black}In this work we present the solution for a rotating Kerr black hole in the weak-field limit under
the radiation gauge proposed by Chen and Zhu [Phys. Rev. D83, 061501(R) (2011)], with which the two physical components of the gravitational wave can be picked out exactly.}
%in the weak-field limit [{\it Phys. Rev.} D83, 061501(R) (2011)]. In this work we derive the solution for Kerr black hole in the weak-field limit under this radiation gauge.
\begin{description}
\item[PACS numbers] 04.20.-q, 04.20.Cv, 04.20.Jb %04.30.-w, 11.15.-q, 04.70.Bw
\end{description}

\end{abstract}

\maketitle

\section{Introduction} \label{sec:intro}

%In general, Einstein field equations do not have enough independent equations to fix the metric due to Bianchi identities. Therefore, we usually need to choose a gauge to obtain a solution to Einstein field equations. Recently, Chen and Zhu propose a true radiation gauge~\cite{ChenZhu2011},
{\color{black}The true radiation gauge proposed by Chen and Zhu~\cite{ChenZhu2011}
\begin{equation}
g^{ij}\Gamma^{\lambda}_{ij}=0~,\label{TG}
\end{equation}
with Greek indices running from $0$ to $3$ and Latin indices running from $1$ to $3$, can pick out exactly the two physical components of the gravitational wave in the weak-field limit, which are shown to give actual gauge invariant physical quantities.}
%In this gauge, the coordinates $x_{\rho}$ obey the Laplace equation
%\begin{equation}
%g^{ij}D_iD_j x^{\rho} = g^{ij}\partial_i \partial_j x^{\rho} + g^{ij} \Gamma^{\lambda}_{ij} \partial_{\lambda} x^{\rho} = 0~.\label{Laplace}
%\end{equation}
%where Greek indices run from $0$ to $3$ and Latin indices run from $1$ to $3$. This gauge can fix the gauge completely and picks out exactly the two physical components of the gravitational field in the weak-field limit.
The detailed discussions on the significance and advantages of this radiation gauge can be found in the original paper by Chen and Zhu~\cite{ChenZhu2011}. In their work, they have given the metric for {\color{black}a} Schwarzschild black {\color{black}hole} in the weak-field limit under the radiation gauge.
Here, we present the weak-field-limit solution for {\color{black}a} Kerr black hole in this radiation gauge.
%It is interesting and also important to study the metric for Kerr black hole in this radiation gauge.

%In In this work we present the solution for Kerr black hole in the weak-field limit under this radiation gauge.

%In our recent work, we have obtained a general solution for Schwarzschild black in the radiation gauge~\cite{LinJiang2016}. In that work, we find that the relation between the radial variable $R$ in the standard form of Schwarzschild metric and the radial one $r$ in the radiation gauge is
%\begin{equation}
%R=r+\frac{3}{2}M~,\label{Rr}
%\end{equation}
%where $M$ is the mass of Schwarzschild black hole, and the gravitational constant and the light speed have been set as $1$.

\section{Derivation} \label{sec:deriv}
We start with the solution for {\color{black}a} Kerr black hole in the Boyer-Lindquist coordinates
~\cite{Boyer1967}, %can be expressed in Boyer-Lindquist coordinates\cite{Boyer1967} in the form:
\begin{eqnarray}
&&  ds^{2} \! = \! - \! \left( \! 1 \! - \! \frac{2m\bar{r}}{\bar{\rho}^{2}}\right)d\bar{t}^{2} \! + \! \frac{\bar{\rho}^{2}}{\bar{\Delta}}d\bar{r}^{2}+\bar{\rho}^{2}d\bar{\theta}^{2} \!  + \! \left(\bar{r}^{2}\!+\!a^{2} \! + \! \frac{2ma^{2}\bar{r}\sin^{2}\bar{\theta}}{\bar{\rho}^{2}}\right)\sin^{2}\bar{\theta} d\bar{\varphi}^{2} \nonumber
   \\
   && ~~~~~~~~~ -  \frac{4ma\bar{r}\sin^{2}\bar{\theta}}{\bar{\rho}^{2}}d\bar{t} d\bar{\varphi}~,\label{BL}
\end{eqnarray}
where the gravitational constant and the light speed have been set as $1$, $m$ and $a$ are the mass and angular momentum per {\color{black}unit} mass of {\color{black}a} Kerr black hole, and $a \le m$ is assumed to avoid naked singularity. {\color{black}Additionally, } $\bar{\rho}^{2}= \bar{r}^{2}+a^{2}\cos^{2}\bar{\theta}$, and $\bar{\Delta}=\bar{r}^{2}+a^{2}-2m\bar{r}$.

Following the similar derivation for the solution of {\color{black}a} Kerr-Newman black hole in the harmonic-coordinate conditions~\cite{LinJiang2014}, we apply the following transformations to Boyer-Lindquist formulation \begin{equation}
  t=\bar{t}~, \quad R=\bar{r}~, \quad \theta=\bar{\theta}~, \quad \varphi=\bar{\varphi}+\int \frac{a}{\bar{\Delta}}d\bar{r}~.
%\varphi=\bar{\varphi}+\int \frac{a}{\bar{\Delta}}d\bar{r}~,
\end{equation}
%\begin{equation}
%  t=\bar{t}~, \quad R=\bar{r}~, \quad \theta=\bar{\theta}~, \quad \varphi=\bar{\varphi}+\int g(\bar{r})d\bar{r}~.
%\end{equation}
and we have
\begin{eqnarray}
&&   ds^{2}  =  A(R,\theta)dt^{2} + 2B(R,\theta)dt~\!dr + 2C(R,\theta)dt~\!d\varphi + D(R,\theta)dR^{2}\nonumber \\
&& ~~~~~~~~~~~~~~~~ + E(R,\theta)d\theta^{2}
+  F(R,\theta)d\varphi^{2}+ 2G(R,\theta)dR~\!d\varphi~,\label{metric2}
\end{eqnarray}
where
\begin{eqnarray}
&&A(R,\theta)=-1+\frac{2mR}{\rho^{2}}~, \nonumber \\
&&B(R,\theta)=\frac{2ma^2R\sin^{2}{\theta}}{\rho^{2}\Delta}~,  \nonumber \\
&&C(R,\theta)=-\frac{2amR\sin^{2}{\theta}}{\rho^{2}}~, \nonumber \\
&&D(R,\theta)=\frac{\rho^{2}}{\Delta}+\left(\frac{a}{\Delta}\right)^{2}\left(R^{2}+a^{2} +\frac{2a^2mR\sin^{2}{\theta}}{\rho^{2}}\right)\sin^{2}{\theta}~, \nonumber \\
&&E(R,\theta)=\rho^{2}~, \nonumber \\
&&F(R,\theta)=\left(R^{2}+a^{2} +\frac{2ma^{2}R\sin^{2}{\theta}}{\rho^{2}}\right)\sin^{2}{\theta}~,\nonumber \\
&& G(R,\theta) = - \frac{a}{\Delta} \left(R^{2}  + a^{2}  + \frac{2ma^{2}R\sin^{2}{\theta}}{\rho^{2}} \!\right)\sin^{2}{\theta}~, \nonumber
\end{eqnarray}
with $\Delta\equiv R^{2}+a^{2}-2mR$, ~and~$\rho^{2}\equiv R^{2}+a^{2}\cos^{2}\theta$~.

%Based on the exact solution for Schwarzschild black hole in the radiation gauge~\cite{LinJiang2016}, we construct a new coordinate system $X_{\mu}$ as follows:
{\color{black}Due to that Kerr metric should reduce to Schwarzschild's} when the spin vanishes ($a=0$), and {\color{black}that} we have obtained the radial transformation from the standard form of Schwarzschild metric to the solution in the radiation gauge~\cite{LinJiang2016}, which is $r = R-\frac{3}{2}m$, we {\color{black}can} construct a new coordinate system $X_{\mu}$ as follows:
\begin{eqnarray}
    && X_{0}=t~, \nonumber\\
    && X_{1}=\sqrt{r^2+a^{2}}\cos\Phi\sin\theta~, \nonumber\\
    && X_{2}=\sqrt{r^2+a^{2}}\sin\Phi\sin\theta~, \nonumber\\
    && X_{3}=r\cos\theta~,\nonumber
\end{eqnarray}\label{HarmonicCoordinates2}
with $r= R-\frac{3}{2}m$, and $\Phi=\varphi+\arctan \frac{a}{r}$. After tedious calculations, we can obtain the solution for {\color{black}a} Kerr black hole in the new system ($t, X_1, X_2, X_3$) as follow
\begin{eqnarray}
&& ds^2=-dt^2 +\frac{\left(r+\frac{3}{2}m\right)^2+\frac{a^2 X_3^2}{r^2}}{\big(r^2+\frac{a^2 X_3^2}{r^2}\big)^2} \bigg[\frac{r^2\left(\bm{X} \!\cdot\! \bm{dX}\!+\!\frac{a^2}{r^2} X_3dX_3\right)^2}{r^2+mr-\frac{3}{4}m^2+a^2}\bigg. \bigg. +\frac{X_3^2 \left(\bm{X} \!\cdot\! \bm{dX}\!-\!\frac{r^2}{X_3^2} X_3dX_3\right)^2}{ \left(r^2-X_3^2\right)}\bigg]\nonumber \\
&&~~~~~~ ~~~~ +\frac{\left(r+\frac{3}{2}m\right)^2+a^2}{r^2+a^2} \!\bigg[\frac{a m \left(r-\frac{3}{4}m\right) \left(\bm{X} \!\cdot\! \bm{dX}\!+\!\frac{a^2}{r^2} X_3dX_3\right)}{\left(r^2+mr-\frac{3}{4}m^2+a^2\right) \left(r^2+\frac{a^2 X_3^2}{r^2}\right)}\bigg.\bigg. -\frac{r (X_2dX_1\!-\!X_1dX_2)}{r^2-X_3^2}\bigg]^2 \nonumber \\
&& \!+\frac{2 m \left(r\!+\!\frac{3}{2}m\right)}{\left(r\!+\!\frac{3}{2}m\right)^2\!\!+\!\frac{a^2 X_3^2}{r^2}}\! \bigg[\frac{a (X_2dX_1\!-\!X_1dX_2 )}{r^2\!+\!a^2}\!-\!\frac{a^2 m \left(r\!-\!\frac{3}{4}m\right) \!\left(r^2\!-\!X_3^2\right) \!\left(\!\bm{X} \!\cdot\! \bm{dX}\!+\!\frac{a^2}{r^2} X_3dX_3\!\right)}{ r\left(r^2\!+\!mr\!-\!\frac{3}{4}m^2\!+\!a^2\right) \!\left(r^2\!+\!a^2\right) \!\left(r^2\!+\!\frac{a^2 X_3^2}{r^2}\right)}\!+\!dt\bigg]^{\!2},\nonumber \\&& ~~
\label{ExactMetric}
\end{eqnarray}
where $\bm{X}\equiv (X_1,~X_2,~X_3)$, ~~$\bm{X} \!\cdot\! d\bm{X}\equiv X_1dX_1 + X_2dX_2 +X_3dX_3$. The relation between the components of $\bm{X}$ and $r$ is $\frac{X_1^2+X_2^2}{r^2+a^2}+\frac{X_3^2}{r^2}=1$~.

Now we demonstrate that this metric satisfies the radiation gauge in the weak-field limit. For $r \rightarrow \infty$, we can expand the metric in powers of the inverse of $r$. Dropping all the terms with the order of $r^{-3}$ and higher, we can rewrite Eq.~\eqref{ExactMetric} as follow
\begin{eqnarray}
&&\hskip -0.5cm ds^2  = - \left(1-\frac{2m}{r}+\frac{3m^2}{r^2}\right)dt^2 + \frac{4 a m (X_2dX_1\!-\!X_1dX_2 )dt}{r^3} \nonumber \\
&&\hskip -0.5cm ~~~~~~~~~~+\left(1+\frac{3m}{r}+\frac{9m^2}{4r^2}\right) d\bm{X}^2 - \left(\frac{m}{r} +\frac{5m^2}{4r^2}\right)\frac{(\bm{X} \!\cdot\! d\bm{X})^2}{r^2}~,\label{Metric_PN15}
\end{eqnarray}
%\begin{equation}
%ds^2  = - \left(1-\frac{2m}{R}\right)dt^2 + \left(1+\frac{3m}{R}\right) d\bm{X}^2 - \frac{m}{R} \frac{(\bm{X} \!\cdot\! d\bm{X})^2}{R^2}+\frac{4 m a (X_2dX_1\!-\!X_1dX_2 )dt}{R^3}~.\label{Metric_PN1}
%\end{equation}
where $r^2 ={\color{black}\bm{X}\!\cdot\!\bm{X}}$~. Substituting this metric into $g^{ij}\Gamma^{\lambda}_{ij}$, we can obtain
\begin{eqnarray}
&&\hskip -0.5cm  g^{ij}\Gamma^{0}_{ij}=0~,\\
&&\hskip -0.5cm g^{ij}\Gamma^{k}_{ij}=-\left[\!\frac{m^3}{4 r^3}\!-\!\frac{3 m^4}{2 r^4}+\frac{87 m^5}{16 r^5}+\frac{2 a^2 m^3}{r^5}\!\left(1-\frac{X_3^2}{r^2}\right)\right]\frac{X_k}{r^2}+O\left(\frac{a^pm^q}{r^{p+q+1}}\right),
\end{eqnarray}
where $p\!+\!q \ge 6$. Here, in order to show the contribution of $a$, we keep the terms up to the order of $\frac{a^pm^q}{r^{p+q+1}}$ with $p\!+\!q\!=\!5$. It can be seen clearly that $g^{ij}\Gamma^{k}_{ij}$ goes to $0$ in the limit of weak field.
%Therefore, Eq.~\eqref{Metric_PN15} satisfies the radiation gauge Eq.~\eqref{TG} in the limit of weak field.
Finally, we calculate the corresponding Ricci tensor for Eq.~\eqref{Metric_PN15} and achieve
\begin{eqnarray}
&& R_{00}=-\frac{27 m^3}{2 r^5}+O\left(\frac{a^pm^q }{r^{p+q+2}}\right)~~,\\
&& R_{0i}=\frac{6 am^2}{r^5} \left(\frac{\epsilon_{ij3} X_j}{r}\right) +O\left(\frac{a^pm^q }{r^{p+q+2}}\right)~, \\%-
%&& R_{01}=\frac{6a m^2}{R^5} \frac{X_2}{R}+O\left(\frac{a^pm^q }{R^{p+q+2}}\right)~, \\%\left(6+\frac{40 m}{R}-\frac{51m^2}{2 R^2}\right)~, \\
%&& R_{ij}=-\frac{7m^3}{R^5}\delta_{ij}  + \frac{65m^3}{2R^5} \left(\frac{X_i X_j}{R^2}\right)+O\left(\frac{a^pm^q }{R^{p+q+2}}\right)~, \\%
&& R_{ij}=-\frac{7m^3}{r^5}\left(\delta_{ij}  - \frac{65}{14} \frac{X_i X_j}{r^2}\right)+O\left(\frac{a^pm^q }{r^{p+q+2}}\right)~,
\end{eqnarray}
where $p\!+\!q\ge 4$, and $\epsilon_{ijk}$ is Levi-Civita symbol. It can be seen that all components of Ricci tensor go to $0$ in the weak-field limit. Therefore, Eq.~\eqref{Metric_PN15} is the weak-field solution to Einstein field equations for {\color{black}a} Kerr black hole in the radiation gauge. It is worth emphasizing that Eq.~\eqref{ExactMetric} is an exact solution for {\color{black}a} Kerr black hole, but only in the weak-field limit does it satisfy the radiation gauge.
%\begin{eqnarray}
%\lim_{R \rightarrow \infty} g^{ij}\Gamma^{\lambda}_{ij} = 0~.
%\end{eqnarray}
%It is easy to check that this solution reduces to the first-order result given in the work~\cite{ChenZhu2011} in the weak-field limit.

%\begin{eqnarray}
%&& ds^2  = - \left(1-\frac{2m}{R}+\frac{3m^2}{R^2}\right)dt^2 + \left(1+\frac{3m}{R}+\frac{9m^2}{4R^2}\right) d\bm{X}^2 - \left(\frac{m}{R} +\frac{5m^2}{4R^2}\right)\frac{(\bm{X} \!\cdot\! d\bm{X})^2}{R^2}\nonumber \\
%&& ~~~~~~~~~~+\frac{4 m a (X_2dX_1\!-\!X_1dX_2 )dt}{R^3}~.\label{Metric_RG}
%%&& g_{ij}=\left[\frac{1+\frac{3 G M}{2 r}}{1-\frac{G M}{2 r}}-\left(1+\frac{3 G M}{2 r}\right)^2\right] \frac{\left(x_i x_j\right)}{r^2}+ \left(1+\frac{3 G M}{2 r}\right)^2\delta _{ij} ~.
%\end{eqnarray}

%
%%The potential application of this metric
%\begin{equation}
%ds^2 \! = \!- \frac{1\!-\!\frac{1}{2} G M/r}{1\!+\! \frac{3}{2} G M/r} dt^2 \!+\! \left(1\!+\!\frac{3}{2} \frac{G M}{r}\right)^{\!2} \! d\bm{x}^2 \!+\! \left[\frac{1\!+\!\frac{3}{2} G M/r}{1\!-\!\frac{1}{2}G M/r} \!-\! \left(\!1\!+\!\frac{3}{2}\frac{G M}{r}\right)^{\!2} \right]\!\frac{(\bm{x} \!\cdot\! d\bm{x})^2}{r^2}~.\label{Metric_RG}
%%&& g_{ij}=\left[\frac{1+\frac{3 G M}{2 r}}{1-\frac{G M}{2 r}}-\left(1+\frac{3 G M}{2 r}\right)^2\right] \frac{\left(x_i x_j\right)}{r^2}+ \left(1+\frac{3 G M}{2 r}\right)^2\delta _{ij} ~.
%\end{equation}

\section{Summary}
In summary, we have derived the weak-field-limit solution for {\color{black}a} Kerr black hole in the radiation gauge. %, which is given in Eq.~(\ref{Metric_PN15}). %This work can be regarded as a preliminary step in understanding the highly complicated nonlinear gravity. On the other hand, the post-Newtonian dynamics has been calculated based on the harmonic coordinate conditions.
This solution may be useful in exploring the gravitational physics for {\color{black}the} Kerr black hole{\color{black}, especially for calculating the energy of gravitational wave in the Kerr spacetime.} %post-Newtonian dynamics and gravitational-wave radiation

\begin{acknowledgements}

{\color{black}We thank the reviewer for providing helpful comments and suggestions to improve the quality of this paper.} This work was supported in part by the National Natural Science Foundation of China (Grant No. 11647314) and Educational Commission of Hunan Province of China (Grant No. 16C1367).

\end{acknowledgements}



\end{document}